\documentstyle[11pt,newpasp,twoside]{article}
\def\edcomment#1{\iffalse\marginpar{\raggedright\sl#1\/}\else\relax\fi}
\reversemarginpar

\begin{document}
\title{Interactions between Pulsars, Pulsar  Nebulae and Supernova Remnants}
 \author{Roger A. Chevalier}
\affil{Department of Astronomy, University of Virginia, P.O. Box 3818,
Charlottesville, VA 22903, USA}

\begin{abstract}

The  Crab Nebula is likely to be expanding into freely expanding supernova
ejecta, although the energy in the ejecta may be less than is typical
for a Type II supernova.
Pulsar nebulae much younger than the Crab have not been found and
could have different properties.
The search for such nebulae through ultraviolet/optical line emission
in core collapse supernovae, or through their X-ray emission (which could
show strong absorption) is warranted.
Neutron stars have now been found in many young supernova remnants.
There is not a clear link between neutron star and remnant type, although
there is an indication that normal pulsars avoid the O-rich remnants.
In the later phases of a supernova remnant, the pulsar wind nebula is
crushed by the reverse shock front.
Recent simulations show that this process is unstable, which can lead
to mixing of the thermal and relativistic gases, and that the pulsar
nebula is easily displaced from the pulsar, which can explain the
position of the Vela pulsar relative to the Vela X radio nebula.

\end{abstract}

\section{Introduction}

Pulsars are thought to be born in the core collapse and explosion of
massive stars, so that the initial evolution and observability of
the PWNe (pulsar wind nebulae) depends on the properties of the supernova
and the supernova remnant evolution.
Several phases of evolution are expected (Pacini \& Salvati 1973; Reynolds
\& Chevalier 1984; Chevalier 1998).
Initially the radiation from the PWN is absorbed by the surrounding supernova
gas; high energy photons may be reprocessed and appear at a lower energy.
After $\ga$10's of years, 
the supernova  becomes optically thin and the PWN radiation
directly escapes.
The Crab Nebula may be in this phase of evolution.
After $\sim 10^3$ years for a pulsar like that in the Crab,
the spindown power from the pulsar decreases and the PWN fades.
After $\sim 10^4$ years, the reverse shock wave from the supernova remnant
moves to the center of the remnant, crushing the PWN and leading
to an increase in brightness of the nebula.

The discussion of PWNe has typically centered on objects like the Crab and
its pulsar, which is estimated to have an initial spin period $P_0
\approx 19$ ms and a magnetic field of $5\times 10^{12}$ G.
It is thought to belong to the main (not binary related) group of 
radio pulsars, which appear to have a fairly narrow distribution of
initial magnetic field strengths $\sim 3\times 10^{12}$ G (Stollman 1987).
The initial rotation rates of these pulsars remain uncertain;
studies of the pulsar population are influenced by selection
effects (e.g., Emmering \& Chevalier 1989).
In recent years there has been increasing evidence for a class of
strongly magnetized neutron stars, or magnetars, with
$B\ga  10^{14}$ G (Thompson \& Duncan 1995).
Magnetars are thought to be associated with soft gamma-ray repeaters
and possibly with AXPs (anomalous X-ray pulsars), both of which appear
to contain slowly rotating neutron stars.
Although the youth of AXPs is indicated by their association with supernova
remnants, there is no evidence for PWNe.
Finally there is a class of ``quiet'' neutron stars that have been found
as weak X-ray sources in young supernova remnants, such as Cas A 
(Tananbaum 1999) and
Puppis A (Petre, Becker,  \& Winkler 1996).
Although the relative formation rates of the normal radio pulsars, magnetars,
and quiet neutron stars
 is not known, the likely youth of the objects in the latter
two categories indicates that all three groups make a significant contribution
to the total rate of massive star core collapses.

Here, I discuss four topics related to neutron stars and their supernova
remnants.
The first is the quest for young PWNe, considerably younger than
the Crab Nebula.
Next, I consider the status of the Crab Nebula and its surroundings.
The third topic concerns the possible relation between a neutron star and
its surrounding supernova, given the growing number of such associations.
Finally, I review recent investigations of the effect of a reverse shock
front on a PWN.

\section{Searching for Young Pulsar Nebulae}

The Crab pulsar, with an age of 947 years, was until recently the youngest
known pulsar.
A newly discovered pulsar in the remnant Kes 75 has a characteristic age
of only 723 years (Gotthelf et al. 2000) and so may be younger.
However, the search for very young pulsars (age $<100$ years) continues.
The discovery of such objects is important for understanding the early
evolution of pulsars.
In their early phases, pulsars are expected to interact
with their surrounding supernovae.
This occurs both through the dynamical interaction of the pulsar bubble
with the supernova and the interaction of the radiation with the 
surrounding supernova gas.
If the pulsar injects energy at a constant rate $\dot E$ in a uniformly expanding
central region of approximately constant density $\rho_c$,
the bubble radius increases as (Chevalier 1977)
\begin{equation}
R=\left({125\over 132\pi}{\dot E\over \rho_c t^3}\right)^{1/5}t^{6/5}
\end{equation}
where $\rho_c t^3$ is a constant because of the uniform expansion.
The shock front generated by the pulsar bubble is expected to initially
be radiative, with a total power $0.015 \dot E$.
The initial spin-down power of a pulsar in the vacuum dipole
approximation can be expressed as
\begin{equation}
\dot E_0=1.6\times 10^{39} \left(B_p \over 3\times 10^{12}{\rm~G}\right)^2
\left(R_{ns}\over {12\rm~km}\right)^6
\left(P_0\over 20{\rm~ms}\right)^{-4}\rm ~erg~s^{-1},
\end{equation}
where $B_p$ is the dipole magnetic field at the pole of the star,
$R_{ns}$ is the neutron star radius, and $P_0$ is the initial period.
A dipole magnetic field perpendicular to the rotation axis has been assumed.
The  velocity of the shell at the bubble boundary is given by
\begin{equation}
V=470\left( \dot E\over 10^{39} \rm ~erg~s^{-1}\right)^{1/5}
\left(\rho_c t^3\over 10^9 {\rm~g~s^3~cm^{-3}}\right)^{-1/5}
\left(t\over 10 {\rm~yr}\right)^{1/5}\rm ~km~s^{-1},
\end{equation}
where the reference value of $\rho_c t^3$ is based on models for SN 1987A.

The radiation can be supplemented by synchrotron radiation from the
pulsar bubble.
The radiation from a pulsar bubble like the Crab Nebula appears to
be from relativistic electrons and positrons that have been accelerated
in a shock front which begins the deceleration of the pulsar wind
(Rees \& Gunn 1974; Kennel \& Coroniti 1984a, 1984b).
Kennel \& Coroniti (1984b) find that the magnetization parameter
$\sigma$ of the pulsar wind (ratio of Poynting flux to kinetic energy flux) must be
$\sim 0.003$ to fit the Crab parameters.
However, when $\sigma\la v/c$ flow solutions with a shock front are
no longer possible; the shock wave moves in toward the position of
the pulsar (Kennel \& Coroniti 1984a; Emmering \& Chevalier 1987).
In the case of the Crab, $v\approx 2000\rm ~km~s^{-1}$, so that $\sigma\la 6.7\times 10^{-3}$
is needed.
The value indicated for the Crab is only slightly smaller than the limit.
According to eqn. (1) with
$V\propto t^{1/5}$, the current value of $\sigma$ was below the limit
at an age of $\sim 20$ years,
which shows that a young pulsar nebula may have different properties than
the Crab.

As $\sigma$ approaches $v/c$ in nebular models, the fact that the shock front
moves in to the central pulsar suggests that the rotating pulsar is in
direct causal contact with the surrounding medium (Emmering \& Chevalier 1987;
Lyutikov 2001).
There is no supersonic wind zone.
The pulsar spindown law is then not given by eqn. (2), but is determined
by the torque acting on the pulsar from the surrounding supernova gas
through the magnetic field.
The pulsar may do work on the surrounding medium, leading to some radiation, but the
normal nebular synchrotron radiation does not occur.

If a pulsar wind has a low magnetization parameter or the pulsar is
powerful so it can drive a fast surrounding shell, it is possible that
even a young pulsar can have a surrounding pulsar nebula with
synchrotron emission from shock accelerated particles.
Chevalier \& Fransson (1992) investigated the emission from a supernova
that contains such a pulsar nebula.
During the first 10 years, the emission is absorbed by the core gas
around the pulsar nebula and re-radiated at optical and ultraviolet
wavelengths.
The line widths depend on the pulsar power, but a typical width is
$\sim 1000\rm ~km~s^{-1}$.
If the core material is oxygen rich, the expected strong lines are
[O III] $\lambda\lambda$4959,5007, O I $\lambda$7774, 
Mg I] $\lambda$4571, and various ultraviolet O lines.
A characteristic property of the line widths is that they should
broaden with time because as the supernova density decreases the ionized region
becomes broader.

The best tests for the presence of a young pulsar are in the nearest
core collapse supernovae, SN 1987A in the LMC  (Large Magellanic Cloud)
and SN 1993J in M81.
The ultraviolet/optical spectrum of SN 1987A is well explained by power input from
$\sim 10^{-4}~M_\odot$ of $^{44}$Ti; any other power source is at
a level $\la 10^{36}\rm ~erg~s^{-1}$ (Chugai et al. 1997).
If a quiet neutron star is present in SN 1987A, accretion onto the neutron
star is expected, giving an Eddington limit luminosity.
An accretion rate of at least the electron scattering Eddington rate
is expected for $\sim 200$ years (Chevalier 1989), but the observed luminosity
limit is considerably lower than the expected $2\times 10^{38}\rm ~erg~s^{-1}$.
One possibility is that a high Fe opacity effectively decreases the
Eddington limit; a photon driven wind can then clear the region around
a central neutron star on a timescale of $\sim 1$ yr (Fryer, Colgate, \& Pinto  1999).
Better limits on a central source of emission will become available
because SN 1987A can now be spatially resolved at 
optical (Pun et al. 2001), X-ray (Park et al. 2001a), and
radio  (Manchester et al. 2001) wavelengths; in no case is there any sign
of central emission.

The case of SN 1993J is complicated by the fact that it is undergoing
strong circumstellar interaction (Matheson et al. 2000).
The lines that are mentioned above as resulting from a pulsar nebula
can also be produced by circumstellar interaction and are observed
in SN 1993J.
Methods of distinguishing between the sources of power are that 
the circumstellar interaction lines tend to be initially broader and to narrow
with time because of deceleration, whereas the pulsar nebula lines
are initially narrower and broaden with time, as discussed above.

While extragalactic supernovae are excellent sites to search for
very young pulsar nebulae, somewhat older ones might be found by
searching for extragalactic objects like the Crab Nebula.
In the LMC,
 the remnant 0540-69 has a pulsar nebula very much like
the Crab and may the penultimate core collapse supernova in the galaxy before
SN 1987A (Kirshner et al. 1989).
Beyond the LMC, however, pulsar nebulae have not been identified.
At radio wavelengths, Reynolds \& Fix (1987) failed to find PWNe in the galaxy M33.
At X-ray wavelengths, deep surveys of star forming galaxies are
becoming available and it is likely that PWNe are being detected.
For example, Pence et al. (2001) have detected 110 sources in an image of M101
down to a level of $10^{36}\rm ~erg~s^{-1}$ in the 0.5--10 keV band.
The problem is to distinguish the PWNe from binary and other sources.
A strongly absorbed spectrum is possible if absorption by the
surrounding supernova gas is still important.

\section{The Crab and its Surroundings}

A long-standing mystery concerning the Crab Nebula is the
nature of the medium into which it is expanding and which contains
the pulsar nebula.
There has not yet been an observation of an extended interaction
region around the Crab.
One possibility is that it is expanding into the freely expanding ejecta
of SN 1054 (e.g., Chevalier 1977).
Another is that the surroundings are a relatively slowly moving stellar
wind from mass loss before the supernova (e.g., Nomoto et al. 1982).

There are a number of arguments that appear to favor the ejecta as the surrounding
medium.
In the ejecta case, the outer edge of the Crab is expected to be slowly
accelerating or at least moving at a constant velocity.
The observations of proper motion are consistent with free expansion, even
in the 
extended ``jet'' on the north side of the Crab (Fesen \& Staker 1993).
If the Crab were interacting with a slow wind, deceleration at the outer edge would
be expected.

In the ejecta interaction model, a radiative shock wave with velocity
of  $\sim230\rm ~km~s^{-1}$ is driven into ejecta over much of the lifetime of the Crab
(Chevalier \& Fransson 1992).
The shocked ejecta cool and accumulate in the observed optical filaments.
The filamentary structure can be attributed to the action of the
Rayleigh-Taylor instability.
Jun (1998) has shown in numerical simulations how the instability can give
rise to structures like those observed in the Crab.
Sankrit \& Hester (1997) have found observational evidence that the
radiative shock wave is still present in the Crab, based on the strength
of [O III] lines at the boundary of the Crab.
These lines are difficult to form by photoionization and are more
likely to be due to a shock front.
In their picture, the shock front is radiative only over the middle
part of the Crab, so it is currently making a transition from the
radiative to the nonradiative phase.

A recent observation is the ultraviolet spectrum of the Crab pulsar
in a search for absorption by freely expanding ejecta gas (Sollerman et al. 2000).
An absorption feature in the C IV $\lambda$1550 line was found with a
negative velocity extent to $-2500\rm ~km~s^{-1}$.
This velocity is somewhat beyond the velocity of the filaments in the
Crab Nebula, but not by much.
It does not extend to the high velocities that would be needed if
an explosion energy of $10^{51}$ ergs were in the freely expanding ejecta
with a plausible density profile.
One of the initial arguments for expansion into free ejecta was that
the ejecta could be carrying a normal supernova energy, as opposed to
the low energy that is directly observed in the Crab Nebula (Chevalier 1977).

The overall result then is that the case for interaction with freely 
expanding ejecta is good, but there is no evidence for a normal
supernova energy in the ejecta.
The lower limit found by Sollerman et al. (2000) is only
$\sim 1.5\times 10^{49}$ ergs.
H$\alpha$ emission should provide a means of detecting more extended,
cool ejecta, but only upper limits have been obtained so far
 (Fesen, Shull,
\& Hurford 1997).
Type II supernovae with low expansion velocities have been found,
such as SN 1994W (Sollerman, Cumming, \& Lundqvist 1998) 
and SN 1997D (Turatto et al. 1998),
but it is not clear whether these supernovae have unusually low energies.
%It is possible that SN 1054 belongs to a class of Type II supernovae with
%low expansion velocities and eenergies; an example of an extrangalactic
%supernova in this category is SN 1997D (Benetti et al. 2001).

\section{Neutron Stars and Their Supernovae}

The increasing number of neutron stars discovered in association with young 
supernova remnants gives the possibility of searching for correlations
between core collapse supernovae and their central compact objects.
Table 1 contains a list of core collapse supernovae with ages
$\la 3,000$  years.
The ages of 3C 58, G11.2--0.3, and MSH 15--52 have been put in
parentheses because uncertain identifications with SN 1181, SN 386, and
SN 185, respectively, have been assumed.
%The X-ray luminosity, $L_x$, is that in the 0.2--4 keV band.
The first 10 objects are in the Galaxy, the next 3 are in the LMC,
and the last one is in the SMC.
If there is not direct evidence for a compact object,
the determination of a supernova being of the core collapse variety
is made based on element overabundances in the ejecta or in the
circumstellar medium (CSM).
A controversial object is Kepler's remnant, which has been suggested to have
a massive star origin based on the circumstellar material (Bandiera 1987).
Gerardy \& Fesen (2001) note the similarity between the spectrum of
a circumstellar knot in Kepler and the quasi-stationary flocculi in Cas A.

The nature of the compact source and the X-ray luminosity of the PWN 
(column 3 and 4) are related in that normal pulsars appear to be
capable of creating synchrotron wind nebulae.
The quoted X-ray luminosity, $L_x$,  is in the 0.2--4 keV range.
The pulsar in 3C 58 was discovered only recently (Murray et al. 2001).
The compact source in G292.0+1.8 is  likely to be a normal pulsar, given the
other correspondences with PWNe.

\begin{table*}

%\noindent{Table 1. Supernova Remnants and their Neutron Stars }
\caption{Supernova Remnants and their Neutron Stars }

%\vspace{1cm}

\begin{tabular}{cccccc}
\hline
  Supernova & Age &  Compact  &  PWN $L_x$  & Element & {Refs.\tablenotemark{a}} \\
    Remnant  & (years)  & Source & (ergs s$^{-1}$)    &   overabundances     & \\
\hline
 Cas A   &  320    & weak   &    &  O, `Si,' N(csm)  & 1,2 \\
 Kepler  &  397   &       &      &   N (csm)  &  3 \\
 Kes 75   &   $\sim 700$   &  pulsar   &  $4\times 10^{35}$  &    & 4,5 \\
 3C 58   &   (820)   &  pulsar   &  $2.9\times 10^{34}$  &  N (csm)  & 6,7 \\
 Crab &  947 &  pulsar    & $2.3\times 10^{37}$  &  H poor, He  & 8 \\
 G292.0+1.8 & $\sim 1000$  &  yes & $4\times 10^{34}$  &  O, Ne  & 9,10 \\
 G11.2--0.3  &   (1615)   &  pulsar   &  $4\times 10^{34}$  &    & 11 \\
 MSH 15--52  &   (1816)   &  pulsar   &  $1.5\times 10^{35}$  &    & 12 \\
 RCW 103   & $\sim 2000$   & weak   &    &    &  13   \\
 Puppis A   &  3700  & weak   &    &  O, Ne, N(csm)  & 14,15,16 \\
 SNR 1987A   & 14   &      &   $< 10^{36}$  &   O, etc.  & 17 \\
 0540--69  &  800  &  pulsar  & $1\times 10^{37}$  &  H poor  & 18,19 \\
 N132D  &  3000  &     &       &   O, Ne, C, Mg   & 20  \\
 1E 0102   &   1000--2000  &      &       &    O, Ne, C, Mg   &  20 \\
 \hline
\end{tabular}
\tablenotetext{a}{The references to the data are: 
(1) Chevalier \& Kirshner 1978; (2) Hughes et al. 2000; (3) 
Leibowitz \& Danziger 1983; (4) Blanton \& Helfand 1996; (5)
Gotthelf et al. 2000; (6) Fesen et al. 1988 ; (7) Murray et al. 2001; (8) 
Davidson \& Fesen 1985; (9) Murdin \& Clark 1979; (10) Hughes et al. 2001;
(11) Torii, et al. 1997; (12) Seward et al. 1984b;  (13) Tuohy \& Garmire 1980;
(14) Canizares, \& Winkler 1981;
(15) Winkler \& Kirshner 1985 ; (16) Petre et al. 1996; (17) Kozma \& Fransson
1998; (18) Seward et al. 1984a; (19) Kirshner et al. 1989; (20)
Blair et al. 2000}

\end{table*}

The weak X-ray sources in Cas A, RCW 103,  and Puppis A appear to be qualitatively
different from the pulsars.
In the case of Puppis A, Gaensler et al. (2001) have set strong
limits on the presence of a radio pulsar nebula.
The other 4 objects have neither an observed compact object nor a
PWN.
Three of these are sufficiently distant that a  Cas A type source
or a weak pulsar would not be observable.
Kepler's remnant is closer and is worthy of a careful search for
a compact object.

The overall impression from Table 1 is that remnants with normal
pulsars and nebulae tend not to have O-rich ejecta, whereas the O-rich
young remnants tend not to have  pulsars.
One possible exception is 0540--69 in the LMC.
This remnant has weak H emission, but lines of O, S,  Ar, Ni, and Fe
have relative strengths that are comparable to those observed
in the Crab Nebula (Kirshner et al. 1989).
This remnant is worthy of detailed abundance studies.
A clearer exception is G292.0+1.8.

The implication is that the pulsars are born in stars with
a range of masses.
The ejecta in the Crab Nebula are not O-rich and are consistent with a low mass star,
$\sim 8-10~M_\odot$ (Nomoto et al. 1982).
The presence of O-rich core material in G292.0+1.8 implies an
initial mass $\ga 13~M_\odot$.
%The initial stellar mass at which a signicant O core develops is
%$\sim 14~M_\odot$ (Arnett 1999).
The tendency of the pulsars to avoid O-rich remnants suggests that they
typically have lower mass progenitors, but
another possibility is that the O-rich remnants are ones with a dense
circumstellar medium, so that the ejecta are observable through
their interaction with the surroundings.
Blair et al. (2000) suggested that N132D and 1E 0102 had Wolf-Rayet star
progenitors, and this suggestion has long been made for Cas A (Chevalier 1976).
However, G292.0+1.8 is a case with a pulsar where strong circumstellar
interaction appears to be taking place (Park et al. 2001b).
There is thus not a clear link between neutron stars and their
supernovae, but the data are reaching the point where interesting
questions can be asked.
%The mass loss may be related to the initial stellar mass, but the
%relation between these is poorly understood.
%A more massive, more luminous star is expected to drive a more powerful wind,
%but the radius of the progenitor star is also a factor in that 
%a faster wind (from a smaller star) leads to a lower density circumstellar medium.

\section{The Reverse Shock Wave Interaction}

In the later stage of evolution of a supernova remnant, the reverse
shock front in the ejecta moves back to the center of the remnant,
crushing a central PWN.
If the surrounding density is $\sim 1$ H atom cm$^{-3}$, this occurs
at an age $\sim 10^4$ years.
Reynolds \& Chevalier (1984) discussed the effects of the crushing on
increasing the synchrotron luminosity of a PWN.
There have recently been hydrodynamic studies of the crushing process.

Van der Swaluw et al. (2001) et al. carried out one dimensional 
simulations of the process using a one fluid code.
The expansion of the PWN inside the supernova was approximately as described
by the analytical model of Reynolds \& Chevalier (1984) during the pre-
and post-reverse shock phases, but there were considerable transient
waves during the transition.
During the late phase, the radius of the PWN increases as $t^{0.3}$,
which is close to the expansion rate of an adiabatic blast wave $t^{0.4}$.
The ratio of the radius of a PWN to the radius of the outer shock front
can thus be related to the energy injected by the pulsar, with only a weak
dependence on age.
Van der Swaluw \& Wu (2001) use this property to obtain estimates 
of the total energy deposited into the PWN nebula for a number
of observed cases and thus deduce the
initial periods of the pulsars.
The result is that the deduced periods are $\ga 10$'s of seconds,
i.e. relatively long.
However, radiative losses from the PWN can allow a larger initial energy.
In addition, a large energy deposited early in the evolution is primarily
lost to adiabatic expansion, so that this method is not very sensitive
to the initial period (Blondin, Chevalier, \& Frierson 2001).

Blondin et al. (2001) carried out two dimensional, two fluid simulations
of the late evolution of a PWN and its surrounding supernova remnant.
The PWN was treated as a $\gamma=4/3$ gas with a high sound speed.
These simulations showed that the crushing and re-expansion of the
PWN are subject to hydrodynamic instabilities so that the thermal and
nonthermal gases become mixed.
The simulations did not include a magnetic field; the formation of
magnetic filaments is likely to occur.
In addition, the reverse shock front is unlikely to be spherically
symmetric, so that the PWN can be pushed to the side relative to
its parent pulsar.
This effect is most likely to be observed at radio wavelengths because
of the large ages of the radio emitting electrons, and can explain
the off-center position of the Vela pulsar relative to the Vela X
nebula.
Off center radio pulsar nebulae have also been found in a number of other
sources.

\section{Discussion}

%The increase in the number of neutron stars in supernova remnants
Of the various phases expected for pulsar nebula evolution, the
earliest phase when nebular emission is absorbed by the
surrounding supernova is least understood because of a lack
of observations.
The next phase when the PWN can be observed directly and is
expanding into freely expanding supernova ejecta is presumably
observed in many cases, although clear proof of this phase has
been difficult to obtain even in the case of the Crab Nebula.
Most of the remnants listed in Table 1, with ages $\la 3000$ yr,
are likely to be in this phase.
Continued study of these sources, especially analysis of the
composition from X-ray emission, is warranted in order to
search for correlations between neutron stars and their
progenitor stars.
In the late, post-reverse shock phase, a pulsar nebula is
easily distorted and displaced by the thermal gas.
Careful studies of the boundaries of these nebulae should
reveal the nature of the interaction.

\acknowledgments
This work was supported in part by NASA grant NAG5-8130.

\end{document}